\documentclass[aps,pra,reprint,superscriptaddress,floatfix,showpacs]{revtex4-1}
\usepackage{blindtext}
\usepackage[utf8]{inputenc}
\usepackage[T1]{fontenc}
\usepackage{bm}
\usepackage{dcolumn}
\usepackage{graphicx}
\usepackage{subfigure}
\usepackage{xcolor}
\begin{document}
\title{An electroplating-based plasmonic platform for giant emission enhancement in monolayer semiconductors}
\author{Abhay Anand V S}
\author{Mihir Kumar Sahoo}
\affiliation{Laboratory of Optics of Quantum Materials (LOQM), Department of Physics, IIT Bombay, Mumbai, 400076, Maharashtra, India}
\author{Faiha Mujeeb}
\affiliation{Department of Physics, IIT Bombay, Mumbai, 400076, Maharashtra, India}
\author{Abin Varghese}
\affiliation{Department of Electrical Engineering, IIT Bombay, Mumbai, 400076, Maharashtra, India}
\author{Subhabrata Dhar}
\affiliation{Department of Physics, IIT Bombay, Mumbai, 400076, Maharashtra, India}
\author{Saurabh Lodha}
\affiliation{Department of Electrical Engineering, IIT Bombay, Mumbai, 400076, Maharashtra, India}
\author{Anshuman Kumar}
\email{anshuman.kumar@iitb.ac.in}
\affiliation{Laboratory of Optics of Quantum Materials (LOQM), Department of Physics, IIT Bombay, Mumbai, 400076, Maharashtra, India}

%\keywords{Electroplating; Gold nanorod; Hyperbolic metamaterial; Plasmonics; Optical biosensor}

\begin{abstract}
Two dimensional semiconductors have attracted considerable attention owing to their exceptional electronic and optical characteristics. However, their practical application has been hindered by the limited light absorption resulting from their atomically thin thickness and low quantum yield. A highly effective approach to manipulate optical properties and address these limitations is integrating subwavelength plasmonic nanostructures with these monolayers. In this study, we employed electron beam lithography and electroplating technique to fabricate a gold nanodisc (AuND) array capable of enhancing the photoluminescence (PL) of monolayer MoS$_2$ giantly. Monolayer MoS$_2$ placed on the top of the AuND array yields up to 150-fold PL enhancement compared to that on a gold film. We explain our experimental findings based on electromagnetic simulations.
\end{abstract}
%\begin{document}

%\flushbottom
%\maketitle
%\thispagestyle{empty}

\maketitle
\section{Introduction}
Transition metal dichalcogenides (TMDCs) like  MoS$_2$ are 2D semiconductors which attracted enormous attention in the past decade due to their loftier optical and electronic properties\cite{mak2010atomically,wang2012electronics}. 
Monolayer TMDCs are direct bandgap semiconductors which exhibit promising applications in light-emitting devices\cite{yin2014preparation}, field effect transistors\cite{radisavljevic2011single,yoon2011good}, biosensors\cite{kalantar2016biosensors,sarkar2014mos2}, plasmon exciton coupling\cite{liu2016strong,li2017tailoring,zeng2017highly}, and 
valleytronics\cite{schaibley2016valleytronics,mak2018light,mak2012control}. Monolayer TMDCs possess stable excitons with large binding energy at room temperatures\cite{splendiani2010emerging, zhu2015exciton,xiao2017excitons}, making them a potential candidate for light-emitting applications. The small light absorption caused by the atomically thin thickness and low quantum yield of monolayer TMDC significantly hinders its practical application. Various methods have been adopted to overcome these issues, such as chemical treatment\cite{amani2015near,amani2016high}, carrier doping\cite{mouri2013tunable}, and manipulation of the nearby local photonic environment\cite{akselrod2015leveraging,wen2017room,butun2015enhanced,you2022resonance,shinomiya2022enhanced,pan2022highly}.

Employing subwavelength nanostructures in conjunction with monolayer TMDCs can effectively improve the emission properties of 2D materials\cite{akselrod2015leveraging,wen2017room,butun2015enhanced,you2022resonance,shinomiya2022enhanced,pan2022highly}. Plasmonic nanoantennas can tune the emission of quantum emitters by enhancing excitation rate and quantum efficiency \cite{ringler2008shaping,li2022plasmon}. The advancement of nanofabrication facilities enables to design and fabricate subwavelength plasmonic nanostructures with precise dimensions. Noble metal nanoparticle-based hybrid system\cite{pan2022highly}, and plasmonic nanostructure arrays fabricated via electron beam lithography (EBL) and metallizations have been widely used in the past decade to enhance the emission properties of TMDC\cite{butun2015enhanced,li2017tailoring,akselrod2015leveraging}.

In this work, we introduce a cylindrical AuND array for enhancing the emission properties of monolayer MoS$_2$. We engineered a gold nanodisc with a diameter of 300 nm and an edge-to-edge gap of 100 nm via EBL, followed by electroplating. Although electroplating is an old technique, it is a less explored area for engineering the emission properties of TMDC. To the best of our knowledge, there is no work reported on nano-electroplating-based plasmonic nanostructures for the engineering of the emission properties of TMDC to date. 

In general, to fabricate an  AuND array, EBL, followed by a physical vapour deposition (PVD) method (e.g., sputtering, thermal evaporation, or electron beam evaporation), is used. While the PVD techniques are known for producing smooth and uniform thin films of Au, they come with limitations, such as the requirement of ultra-high vacuum conditions and a relatively long deposition time. The AuND array fabrication using EBL and the PVD technique often requires multilayer spin coating of electron beam resists for an easy lift-off process. However, a multilayer resist usage can form a truncated disc, which makes the wet transfer of TMDC more difficult on top of AuND and increases the risk of breakage. On the other hand, electroplating is a chemical deposition technique in an ambient atmosphere that requires a short duration to deposit Au. In our previous work, we fabricated a cylindrical AuND array using EBL followed by electroplating using a single-layer resist process \cite{sahoo2023electroplating}.

\begin{figure*}
    \centering
    \includegraphics[width=0.95\textwidth]{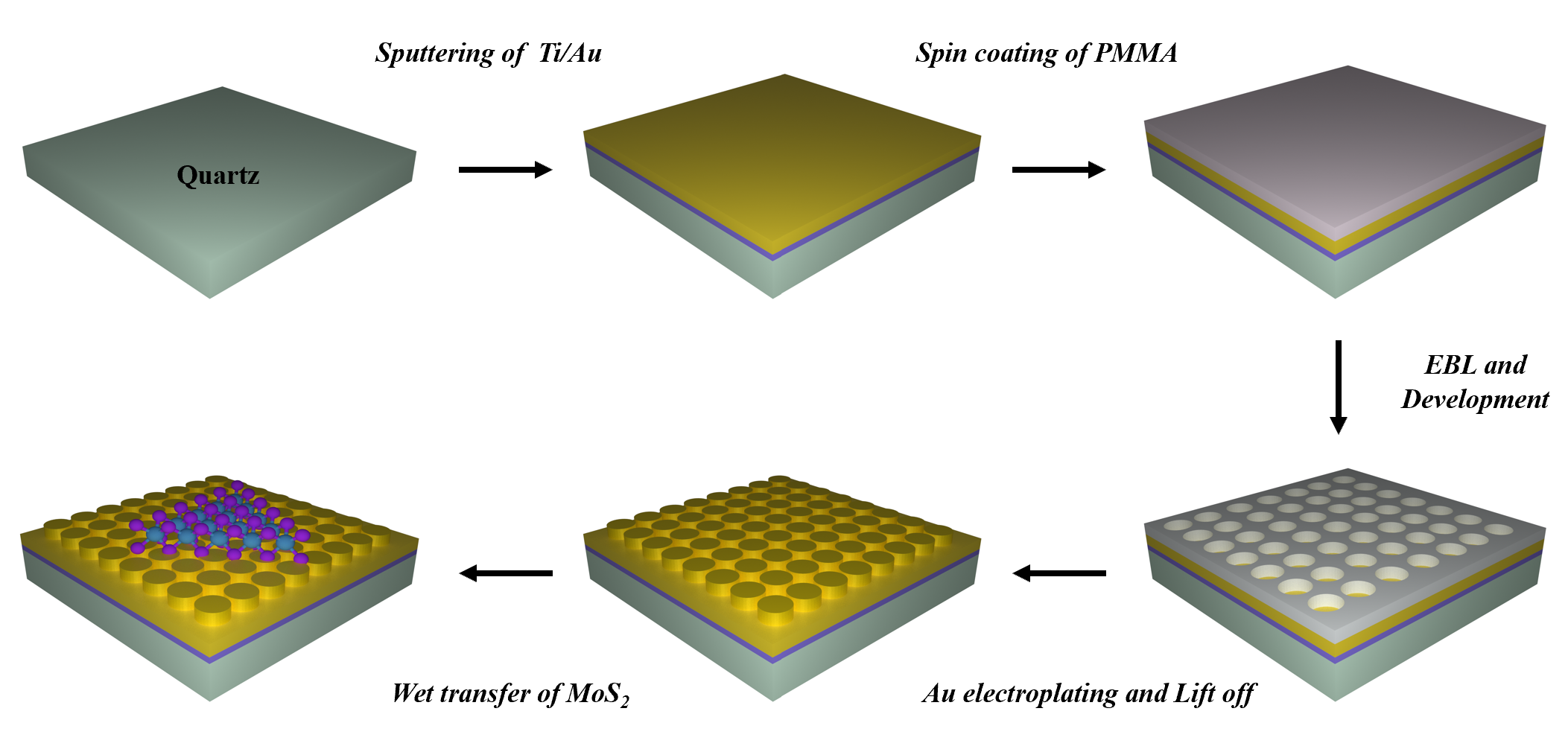}
    \caption{ Process flow of AuND-MoS$_2$ hybrid structure fabrication via electron beam lithography (EBL) and gold (Au) electroplating protocol followed by wet transfer of MoS$_2$.}
    \label{FIG:1}
\end{figure*}

In this study, we introduce an innovative platform based on AuND to enhance the optical properties of monolayer MoS$_2$. The results of our investigation demonstrate that the hybrid structures of AuND-MoS$_2$ display a remarkable 150-fold improvement in photoluminescence (PL) compared to the  MoS$_2$ monolayer on Au film. To further evaluate the efficiency of the AuND array, we performed a PL mapping analysis.  We employed numerical simulations utilizing Lumerical FDTD to elucidate the mechanism behind the enhancement of PL.

\section{Experimental section}
Figure 1 shows the process flow of fabricating the AuND-MoS$_2$ hybrid structure. Details of AuND-MoS$_2$ hybrid structure fabrication and  Lumerical FDTD simulations are described in this section.
\begin{figure*}
    \centering
    \includegraphics[width=1\textwidth]{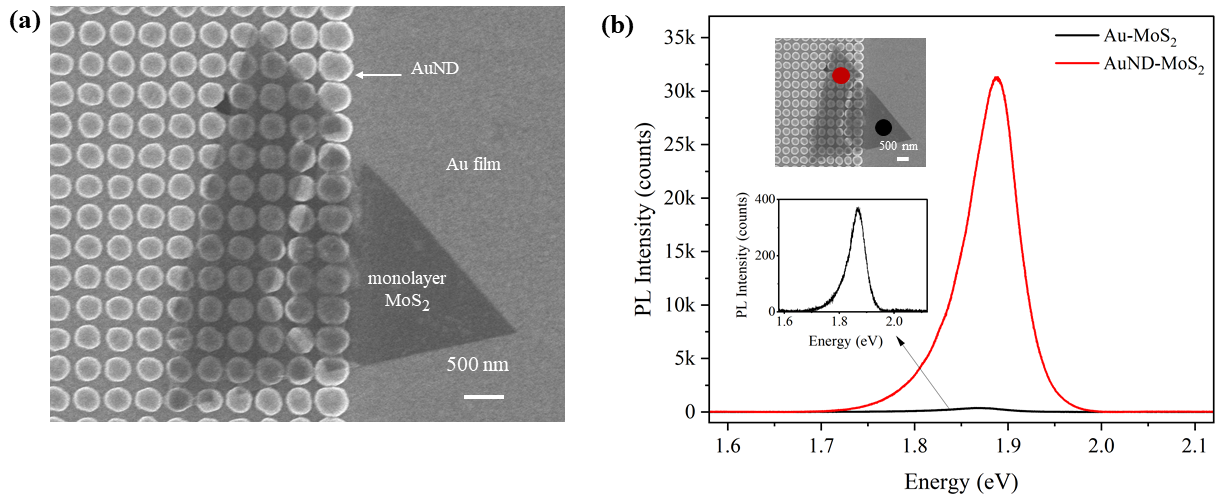}
    \caption{(a) SEM image of AuND-MoS$_2$ hybrid structure. The image shows triangle MoS$_2$ partially sitting on top of the AuND array. (b) PL spectra of monolayer MoS$_2$ on top of gold film and AuND array. PL from AuND-MoS$_2$ hybrid structure is giantly enhanced when compared with Au-MoS$_2$. The top inset image shows the points where the measurements are taken, and the bottom inset shows the PL spectrum of Au-MoS$_2$}
    \label{FIG:2}
\end{figure*}

\subsection{EBL patterning of the AuND array}
  Quartz substrates were cleaned using ultrasonication in trichloroethylene, acetone, and IPA, for five minutes. Piranha cleaning (H$_2$SO$_4$:  H$_2$O$_2$:: 7 : 3) 
 was carried out to remove the residue of sodium and potassium.
The cleaned quartz substrate was sputtered with a five nm-thick adhesive Ti-layer at a 2.1 nm/min rate, followed by a 20 nm  Au film deposited at a 17 nm/min rate, resulting in a transparent and conductive quartz substrate. This conducting seed layer is essential for the electroplating process\cite{sahoo2023electroplating}.

An array of gold nanodiscs is created on the sample using EBL. The nanodiscs have a diameter of 300 nm and are spaced 400 nm from center to center. A layer of PMMA A2 is spin-coated on the sample at 1000 rpm for one minute and then post-baked for five minutes at 180$^{\circ}$C before being exposed to an electron beam. Using a beam aperture of 10$\mu$m, a beam current of 0.026 nA, and a dosage of 400  $\mu$C $cm^{-2}$, an array pattern of AuND is written onto the PMMA layer at 20 kV. The exposed sample developed in a solution of methyl isobutyl ketone (MIBK) and isopropyl alcohol (IPA) in a 3:1 ratio for 35 seconds to develop the pattern, then transferring the sample to IPA solution for 70 seconds to stop further development.

\subsection{Nano-electroplating}

The AuND array was fabricated on an EBL-processed sample through electroplating. In this method, the bottom Au thin film layer serves as a seed layer for the electroplating deposition. The sample functions as a cathode during electroplating, while a platinum mesh is utilized as an anode. Electroplating of Au was conducted by placing the sample (cathode) and platinum (anode) in the gold sulfite solution in a beaker on a hot plate operating at 120℃ with a magnetic stirrer speed of 300 rpm. A hot gold sulfite solution with a magnetic stirrer prevents big chunk formation on the sample. The DC power supply was operated in constant current mode at 0.002 A DC for 45 sec. The sample was promptly removed and submerged in DI water to prevent immersion deposition\cite{sahoo2023electroplating}. 
 The electroplating method for Au deposition depends on the exposed seed-layer area of the sample to the electrolyte (A), the DC current (I), and the deposition time(t). The thickness of the electroplated Au layer can be formulated as given equation 1.

\begin{equation}
    \text { t }(\mu m)=\left(3.5257 \times 10^{-4}\right) \times \frac{I(\mathrm{~mA})}{A\left(\mathrm{~cm}^2\right)} \times \text { Time }(\mathrm{sec})
\end{equation}

The height we obtained for AuND is 180 nm. The height measurements are done by AFM(Supporting Information S1)

\subsection{CVD Growth of Monolayer MoS$_2$}
The CVD technique used to grow monolayer MoS$_2$ films on double-sided polished c-plane sapphire. A microcavity-based growth technique is used for the CVD\cite{mohapatra2016strictly}. Pure molybdenum trioxide (MoO$_3$) and sulfur (S) were used as precursors for growth.  At the center of a two-inch tubular furnace, a ceramic boat was filled with 6 g of MoO$_3$  powder, while at the end of the reactor furnace, a sulfur boat weighing 350 g was placed.
The temperature of the furnace was increased to 760 $^{\circ}$C  at a rate of 15$^{\circ}$C per minute. As the furnace approached 760 $^{\circ}$C, the sulfur zone reached a temperature of approximately 200$^{\circ}$ C. Argon gas was then used at a flow rate of 10 standard cubic centimeters per minute (sccm) to transport sulfur vapours to the MoO$_3$ zone. The furnace was held at 760 $^{\circ}$C  for 8 minutes to allow growth. Further information on the growth parameters and natural cavity arrangement of substrates on top of the MoO$_3$ boat can be found elsewhere\cite{mohapatra2016strictly}.

\subsection{Transfer of Monolayer MoS$_2$ on  AuND array}
A polystyrene (PS) assisted transfer method was used to transfer monolayer MoS$_2$ from sapphire to AuND array\cite{gurarslan2014surface}. The CVD-grown MoS$_2$ on the sapphire substrate was spin-coated with a PS solution and baked at 80-90 $^{\circ}$C for 35 minutes, followed by 120 $^{\circ}$C for 10 minutes,  which helps to build a strong adhesion between MoS$_2$ monolayer and PS layer. A droplet of water was placed on the surface of the baked PS/MoS$_2$/sapphire sample. The sample edges were delicately prodded with a  tweezer to facilitate the smooth penetration of water through the MoS$_2$ film, which will delaminate the PS-MoS$_2$ assembly from the sapphire. PS-MoS$_2$ assembly was gently picked up with a tweezer and transferred onto the AuND array. The hybrid MoS$_2$-AuND structure was baked again at 80-90 $^{\circ}$C for 35 minutes to remove water residues, followed by 120$^{\circ}$C for 10 minutes to remove wrinkles which may happen at the time of transfer.

\subsection{FDTD simulation}

The nearfield intensity profile of AuND is calculated by using the finite difference time domain (FDTD) method using Lumerical FDTD solution software.
In the simulation setup, a single AuND having a 300 nm diameter is positioned at the center of the xy-plane, with the simulation region of 400 nm spanning  in both the x and y directions. To emulate the periodic arrangement of the AuND array, periodic boundary conditions are employed in both the x and y directions. The quartz is defined using the dielectric constant, while the optical dielectric function of titanium and gold are defined using the data given elsewhere\cite{palik1998handbook}. A fine mesh with a resolution of 1 nm is utilized in all three directions (x, y, and z) for the simulation.
\begin{figure*}
    \centering
    \includegraphics[width=0.95\textwidth]{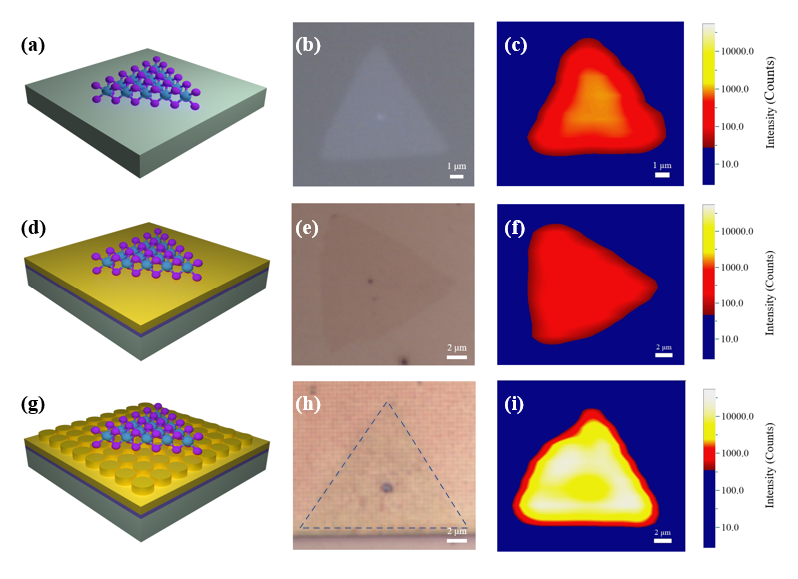}
    \caption{ PL mapping analysis. (a) Schematics of as-grown MoS$_2$ on sapphire, (b) optical image of MoS$_2$ on sapphire, (c) PL mapping of MoS$_2$ on sapphire,(d) Schematics of  MoS$_2$ on Au, (e) Optical image  of  MoS$_2$ on Au, (f) PL mapping of MoS$_2$ on Au, (g) Schematics of  MoS$_2$ on AuND array, (h) Optical image of MoS$_2$ on AuND array (Dashed line shows the outer line of MoS$_2$), (i) PL mapping of MoS$_2$ on AuND array }
    \label{FIG:3}
\end{figure*}

\section{Characterization}

PL and Raman (make: HORIBA/LabRAM HR Evolution) measurements were carried out to study the emission and vibrational properties of MoS$_2$. A 532 nm laser with a power of 42 $\mu$W was used to conduct PL studies. The grating used in the PL study is 600 grooves per mm. A 532 nm laser with a 1.4 mW power took Raman spectra with grating 1800 grooves per mm. The objective is 100x with NA 0.90 for both Raman and PL measurements. SEM (make: Raith GmbH/ 150-Two) and AFM (make: Asylum/Oxford Instruments, MFP3D Origin) characterizations were performed to measure the geometry,  and height of the fabricated AuND array.

\begin{figure*}
    \centering
    \includegraphics[width=0.9\textwidth]{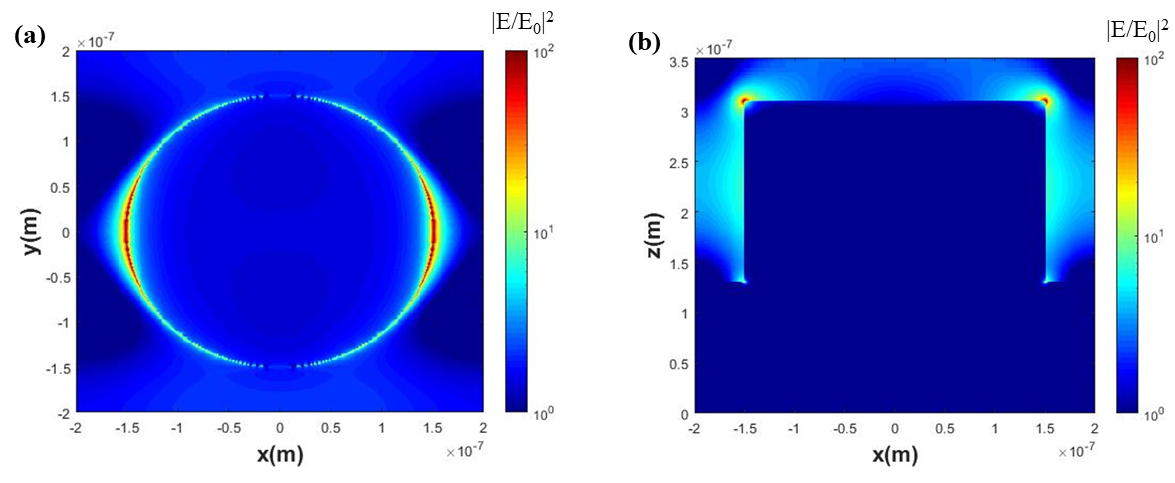}
    \caption{Numerical simulation showing electric field confinement around Au nanodisc at 532 nm excitation (a) Horizontal view of field profile on top of gold nanodisc (b) Cross-sectional view of field profile.}
    \label{FIG:6}
\end{figure*}

\section{Results and Discussion}

Raman spectroscopy verifies the monolayer MoS$_2$. Raman spectra of MoS$_2$ were taken from as-grown CVD sample on sapphire, transferred on Au film and AuND array (Supporting information figure S2). The peak separation between E$_{2g}^1$ and A$_{1g}$ is less than  20 cm$\textsuperscript{-1}$ confirmed the monolayer nature of MoS$_2$\cite{li2012bulk}.

\begin{figure*}
    \centering
    \includegraphics[width=1\linewidth]{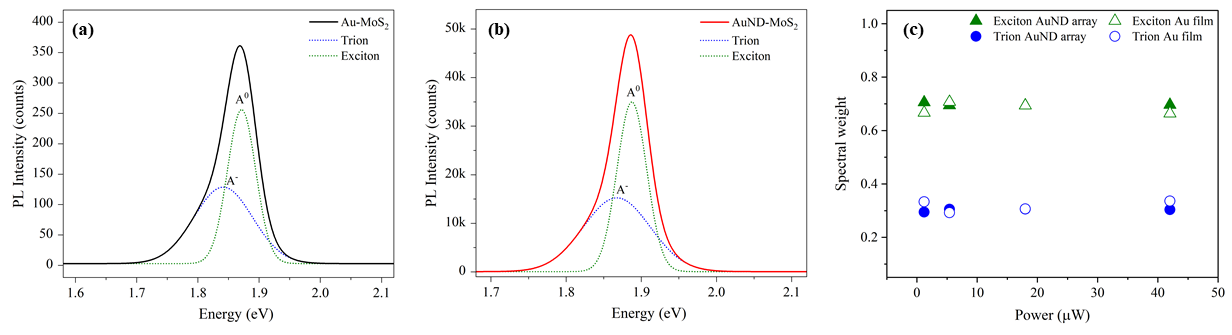}
    \caption{Exciton-trion dynamics in (a) Au-MoS$_2$, (b) AuND-MoS$_2$. (c) Power dependent spectral weight}
    \label{FIG:7}
\end{figure*}
 
\subsection{PL Analysis}
 Figure 2(a) shows the SEM image of AuND-MoS$_2$ hybrid structure. The monolayer MoS$_2$ is partially sitting on the Au film and on the AuND array. Figure 2(b) shows the PL spectrum taken from MoS$_2$ sitting on top of the AuND array and Au thin film. The top inset of Figure 2(b) shows the spots, where measurements were taken. The PL signal of the AuND-MoS$_2$ hybrid structure shows huge enhancement compared to the Au-MoS$_2$. It is important to note that the acquisition parameters remain identical in all measurements. At low excitation intensities, the visibility of the B exciton is typically diminished due to state-filling effects\cite{Plechinger2014}.

 We analysed the PL of MoS$_2$ on as-grown sapphire and transferred MoS$_2$ on Au film and AuND array using PL mapping analysis, as shown in Figure 3. Figure 3(a), 3(b) and 3(c) show the schematic, optical image and PL mapping of as-grown MoS$_2$ on sapphire, respectively. The PL mapping shows an average intensity of 1k-2k counts. Figure 3(d), 3(e) and 3(f) show the schematic, optical image and PL mapping of transferred MoS$_2$ on Au film, respectively. The PL mapping of MoS$_2$ on gold clearly says that PL counts were lower compared to MoS$_2$ on the as-grown sapphire substrate. The reason for PL quenching in Au film is due to charge transfer\cite{bhanu2014photoluminescence}. The average PL count of MoS$_2$ on the Au film was around 250. Figure 3(g), 3(h) and 3(i) show the schematic, optical image and PL mapping of transferred MoS$_2$ on AuND array, respectively. The flakes transferred on top of the AuND array shows a huge enhancement of PL, as shown in Figure 2(b). The doublet observed around 1.87 eV (Figure S2 (c))
is due to the presence of Cr\textsuperscript{3+} luminescence centers from sapphire substrate\cite{dumcenco2015large,kudryashov1999influence}, those signatures were absent in transferred MoS$_2$ on top of Au and AuND.

\subsection{PL enhancement Mechanism}

The intensity of PL from excitons is influenced majorly by the excitation rate and emission rate. The enhancement in excitation is directly proportional to the square of the near-field intensity, as given in equation 2. Plasmon resonance occurring at the excitation wavelength increases the radiative decay rate. 
The nearfield enhancement calculation of plasmonic nanostructures can directly give insights into the higher excitation rate and hence resulting in strong exciton emission.
\cite{wang2016giant, Bharadwaj2007}.
\begin{equation}
    \text{PL enhancement} \propto {{\left|E / E_0\right|}^2}
\end{equation}

A Lumerical FDTD simulation was conducted to investigate the giant PL enhancement mechanism by obtaining field confinement around the AuND array. The plasmonic modes in the AuND array were excited using an x-polarised plane wave source at 532 nm. Our findings revealed a significant hot spot along the circumference of the AuND array, as shown in Figure 4, resulting in an enhancement factor ${\left|E / E_0\right|}^2$ of approximately 100.

\subsection{Exciton-Trion dynamics with excitation power}

To investigate the exciton-trion contribution in PL enhancement, we deconvoluted exciton and trion peaks of the PL spectrum in Figure 2(b). Figure 5(a) shows the PL spectrum of the Au-MoS$_2$ sample, which is fitted using a gaussian profile. Figure 5(b) shows the PL spectrum of the AuND-MoS$_2$ hybrid structure, which is also fitted using a gaussian profile. 

Spectral weight is a quantity that quantifies exciton or trion contribution in PL emission. Spectral weight can be defined as 
\begin{equation}
I_m /\left(I_E+\right.I_T)
\end{equation}
where $I$ is the intensity of PL emission. $E$ represent exciton and $T$ represent trion. $m$ can be $T$ or $E$ according to trion or exciton spectral weight.
Trion recombination pathways are mostly non-radiative\cite{Lien2019,Singh2023}. If exciton spectral weight dominates one over other system, it contributes to PL enhancement. In our Au-MoS$_2$ and AuND-MoS$_2$ hybrid structure, Exciton spectral weight remains around 70 per cent both for Au-MoS$_2$ and AuND-MoS$_2$ system. This is consistent for the power range of 1-40 $\mu$W. This implies exciton spectral weight is not playing any significant role in giant PL enhancement of AuND-MoS$_2$ structure.

\section{Conclusion}

In summary, this study involved the developing of a new platform of a plasmonic array of gold nanodiscs (AuNDs) using electron beam lithography (EBL) and subsequent electroplating. The transferred chemical vapor deposition (CVD)-grown molybdenum disulfide (MoS$_2$) onto the AuND array exhibited a remarkable enhancement of photoluminescence (PL) due to the presence of strong plasmonic hotspots generated by the nanodiscs. We conducted a comparative analysis of the PL intensities between as-grown CVD MoS$_2$, MoS$_2$ transferred onto gold, and MoS$_2$ transferred onto AuNDs. Despite the occurrence of charge transfer in the gold film, the intense plasmonic fields were capable of counteracting the losses and significantly enhancing the PL signal.L.The FDTD simulation showed 100 times electric field enhancement around nanodisc, which is the primary reason of giant PL enhancement.

\newpage
\bibliography{sample}

\begin{thebibliography}{40}%
\makeatletter
\providecommand \@ifxundefined [1]{%
 \@ifx{#1\undefined}
}%
\providecommand \@ifnum [1]{%
 \ifnum #1\expandafter \@firstoftwo
 \else \expandafter \@secondoftwo
 \fi
}%
\providecommand \@ifx [1]{%
 \ifx #1\expandafter \@firstoftwo
 \else \expandafter \@secondoftwo
 \fi
}%
\providecommand \natexlab [1]{#1}%
\providecommand \enquote  [1]{``#1''}%
\providecommand \bibnamefont  [1]{#1}%
\providecommand \bibfnamefont [1]{#1}%
\providecommand \citenamefont [1]{#1}%
\providecommand \href@noop [0]{\@secondoftwo}%
\providecommand \href [0]{\begingroup \@sanitize@url \@href}%
\providecommand \@href[1]{\@@startlink{#1}\@@href}%
\providecommand \@@href[1]{\endgroup#1\@@endlink}%
\providecommand \@sanitize@url [0]{\catcode `\\12\catcode `\$12\catcode
  `\&12\catcode `\#12\catcode `\^12\catcode `\_12\catcode `\%12\relax}%
\providecommand \@@startlink[1]{}%
\providecommand \@@endlink[0]{}%
\providecommand \url  [0]{\begingroup\@sanitize@url \@url }%
\providecommand \@url [1]{\endgroup\@href {#1}{\urlprefix }}%
\providecommand \urlprefix  [0]{URL }%
\providecommand \Eprint [0]{\href }%
\providecommand \doibase [0]{http://dx.doi.org/}%
\providecommand \selectlanguage [0]{\@gobble}%
\providecommand \bibinfo  [0]{\@secondoftwo}%
\providecommand \bibfield  [0]{\@secondoftwo}%
\providecommand \translation [1]{[#1]}%
\providecommand \BibitemOpen [0]{}%
\providecommand \bibitemStop [0]{}%
\providecommand \bibitemNoStop [0]{.\EOS\space}%
\providecommand \EOS [0]{\spacefactor3000\relax}%
\providecommand \BibitemShut  [1]{\csname bibitem#1\endcsname}%
\let\auto@bib@innerbib\@empty
%</preamble>
\bibitem [{\citenamefont {Mak}\ \emph {et~al.}(2010)\citenamefont {Mak},
  \citenamefont {Lee}, \citenamefont {Hone}, \citenamefont {Shan},\ and\
  \citenamefont {Heinz}}]{mak2010atomically}%
  \BibitemOpen
  \bibfield  {author} {\bibinfo {author} {\bibfnamefont {K.~F.}\ \bibnamefont
  {Mak}}, \bibinfo {author} {\bibfnamefont {C.}~\bibnamefont {Lee}}, \bibinfo
  {author} {\bibfnamefont {J.}~\bibnamefont {Hone}}, \bibinfo {author}
  {\bibfnamefont {J.}~\bibnamefont {Shan}}, \ and\ \bibinfo {author}
  {\bibfnamefont {T.~F.}\ \bibnamefont {Heinz}},\ }\href@noop {} {\bibfield
  {journal} {\bibinfo  {journal} {Physical review letters}\ }\textbf {\bibinfo
  {volume} {105}},\ \bibinfo {pages} {136805} (\bibinfo {year}
  {2010})}\BibitemShut {NoStop}%
\bibitem [{\citenamefont {Wang}\ \emph {et~al.}(2012)\citenamefont {Wang},
  \citenamefont {Kalantar-Zadeh}, \citenamefont {Kis}, \citenamefont
  {Coleman},\ and\ \citenamefont {Strano}}]{wang2012electronics}%
  \BibitemOpen
  \bibfield  {author} {\bibinfo {author} {\bibfnamefont {Q.~H.}\ \bibnamefont
  {Wang}}, \bibinfo {author} {\bibfnamefont {K.}~\bibnamefont
  {Kalantar-Zadeh}}, \bibinfo {author} {\bibfnamefont {A.}~\bibnamefont {Kis}},
  \bibinfo {author} {\bibfnamefont {J.~N.}\ \bibnamefont {Coleman}}, \ and\
  \bibinfo {author} {\bibfnamefont {M.~S.}\ \bibnamefont {Strano}},\
  }\href@noop {} {\bibfield  {journal} {\bibinfo  {journal} {Nature
  nanotechnology}\ }\textbf {\bibinfo {volume} {7}},\ \bibinfo {pages} {699}
  (\bibinfo {year} {2012})}\BibitemShut {NoStop}%
\bibitem [{\citenamefont {Yin}\ \emph {et~al.}(2014)\citenamefont {Yin},
  \citenamefont {Zhang}, \citenamefont {Cai}, \citenamefont {Chen},
  \citenamefont {Wong}, \citenamefont {Tay}, \citenamefont {Chai},
  \citenamefont {Wu}, \citenamefont {Zeng}, \citenamefont {Zheng} \emph
  {et~al.}}]{yin2014preparation}%
  \BibitemOpen
  \bibfield  {author} {\bibinfo {author} {\bibfnamefont {Z.}~\bibnamefont
  {Yin}}, \bibinfo {author} {\bibfnamefont {X.}~\bibnamefont {Zhang}}, \bibinfo
  {author} {\bibfnamefont {Y.}~\bibnamefont {Cai}}, \bibinfo {author}
  {\bibfnamefont {J.}~\bibnamefont {Chen}}, \bibinfo {author} {\bibfnamefont
  {J.~I.}\ \bibnamefont {Wong}}, \bibinfo {author} {\bibfnamefont {Y.-Y.}\
  \bibnamefont {Tay}}, \bibinfo {author} {\bibfnamefont {J.}~\bibnamefont
  {Chai}}, \bibinfo {author} {\bibfnamefont {J.}~\bibnamefont {Wu}}, \bibinfo
  {author} {\bibfnamefont {Z.}~\bibnamefont {Zeng}}, \bibinfo {author}
  {\bibfnamefont {B.}~\bibnamefont {Zheng}},  \emph {et~al.},\ }\href@noop {}
  {\bibfield  {journal} {\bibinfo  {journal} {Angewandte Chemie International
  Edition}\ }\textbf {\bibinfo {volume} {53}},\ \bibinfo {pages} {12560}
  (\bibinfo {year} {2014})}\BibitemShut {NoStop}%
\bibitem [{\citenamefont {Radisavljevic}\ \emph {et~al.}(2011)\citenamefont
  {Radisavljevic}, \citenamefont {Radenovic}, \citenamefont {Brivio},
  \citenamefont {Giacometti},\ and\ \citenamefont
  {Kis}}]{radisavljevic2011single}%
  \BibitemOpen
  \bibfield  {author} {\bibinfo {author} {\bibfnamefont {B.}~\bibnamefont
  {Radisavljevic}}, \bibinfo {author} {\bibfnamefont {A.}~\bibnamefont
  {Radenovic}}, \bibinfo {author} {\bibfnamefont {J.}~\bibnamefont {Brivio}},
  \bibinfo {author} {\bibfnamefont {V.}~\bibnamefont {Giacometti}}, \ and\
  \bibinfo {author} {\bibfnamefont {A.}~\bibnamefont {Kis}},\ }\href@noop {}
  {\bibfield  {journal} {\bibinfo  {journal} {Nature nanotechnology}\ }\textbf
  {\bibinfo {volume} {6}},\ \bibinfo {pages} {147} (\bibinfo {year}
  {2011})}\BibitemShut {NoStop}%
\bibitem [{\citenamefont {Yoon}\ \emph {et~al.}(2011)\citenamefont {Yoon},
  \citenamefont {Ganapathi},\ and\ \citenamefont {Salahuddin}}]{yoon2011good}%
  \BibitemOpen
  \bibfield  {author} {\bibinfo {author} {\bibfnamefont {Y.}~\bibnamefont
  {Yoon}}, \bibinfo {author} {\bibfnamefont {K.}~\bibnamefont {Ganapathi}}, \
  and\ \bibinfo {author} {\bibfnamefont {S.}~\bibnamefont {Salahuddin}},\
  }\href@noop {} {\bibfield  {journal} {\bibinfo  {journal} {Nano letters}\
  }\textbf {\bibinfo {volume} {11}},\ \bibinfo {pages} {3768} (\bibinfo {year}
  {2011})}\BibitemShut {NoStop}%
\bibitem [{\citenamefont {Kalantar-zadeh}\ and\ \citenamefont
  {Ou}(2016)}]{kalantar2016biosensors}%
  \BibitemOpen
  \bibfield  {author} {\bibinfo {author} {\bibfnamefont {K.}~\bibnamefont
  {Kalantar-zadeh}}\ and\ \bibinfo {author} {\bibfnamefont {J.~Z.}\
  \bibnamefont {Ou}},\ }\href@noop {} {\bibfield  {journal} {\bibinfo
  {journal} {Acs Sensors}\ }\textbf {\bibinfo {volume} {1}},\ \bibinfo {pages}
  {5} (\bibinfo {year} {2016})}\BibitemShut {NoStop}%
\bibitem [{\citenamefont {Sarkar}\ \emph {et~al.}(2014)\citenamefont {Sarkar},
  \citenamefont {Liu}, \citenamefont {Xie}, \citenamefont {Anselmo},
  \citenamefont {Mitragotri},\ and\ \citenamefont {Banerjee}}]{sarkar2014mos2}%
  \BibitemOpen
  \bibfield  {author} {\bibinfo {author} {\bibfnamefont {D.}~\bibnamefont
  {Sarkar}}, \bibinfo {author} {\bibfnamefont {W.}~\bibnamefont {Liu}},
  \bibinfo {author} {\bibfnamefont {X.}~\bibnamefont {Xie}}, \bibinfo {author}
  {\bibfnamefont {A.~C.}\ \bibnamefont {Anselmo}}, \bibinfo {author}
  {\bibfnamefont {S.}~\bibnamefont {Mitragotri}}, \ and\ \bibinfo {author}
  {\bibfnamefont {K.}~\bibnamefont {Banerjee}},\ }\href@noop {} {\bibfield
  {journal} {\bibinfo  {journal} {ACS nano}\ }\textbf {\bibinfo {volume} {8}},\
  \bibinfo {pages} {3992} (\bibinfo {year} {2014})}\BibitemShut {NoStop}%
\bibitem [{\citenamefont {Liu}\ \emph {et~al.}(2016)\citenamefont {Liu},
  \citenamefont {Lee}, \citenamefont {Naylor}, \citenamefont {Ee},
  \citenamefont {Park}, \citenamefont {Johnson},\ and\ \citenamefont
  {Agarwal}}]{liu2016strong}%
  \BibitemOpen
  \bibfield  {author} {\bibinfo {author} {\bibfnamefont {W.}~\bibnamefont
  {Liu}}, \bibinfo {author} {\bibfnamefont {B.}~\bibnamefont {Lee}}, \bibinfo
  {author} {\bibfnamefont {C.~H.}\ \bibnamefont {Naylor}}, \bibinfo {author}
  {\bibfnamefont {H.-S.}\ \bibnamefont {Ee}}, \bibinfo {author} {\bibfnamefont
  {J.}~\bibnamefont {Park}}, \bibinfo {author} {\bibfnamefont {A.~C.}\
  \bibnamefont {Johnson}}, \ and\ \bibinfo {author} {\bibfnamefont
  {R.}~\bibnamefont {Agarwal}},\ }\href@noop {} {\bibfield  {journal} {\bibinfo
   {journal} {Nano letters}\ }\textbf {\bibinfo {volume} {16}},\ \bibinfo
  {pages} {1262} (\bibinfo {year} {2016})}\BibitemShut {NoStop}%
\bibitem [{\citenamefont {Li}\ \emph {et~al.}(2017)\citenamefont {Li},
  \citenamefont {Li}, \citenamefont {Han}, \citenamefont {Wang}, \citenamefont
  {Yu}, \citenamefont {Tay}, \citenamefont {Liu},\ and\ \citenamefont
  {Fang}}]{li2017tailoring}%
  \BibitemOpen
  \bibfield  {author} {\bibinfo {author} {\bibfnamefont {Z.}~\bibnamefont
  {Li}}, \bibinfo {author} {\bibfnamefont {Y.}~\bibnamefont {Li}}, \bibinfo
  {author} {\bibfnamefont {T.}~\bibnamefont {Han}}, \bibinfo {author}
  {\bibfnamefont {X.}~\bibnamefont {Wang}}, \bibinfo {author} {\bibfnamefont
  {Y.}~\bibnamefont {Yu}}, \bibinfo {author} {\bibfnamefont {B.}~\bibnamefont
  {Tay}}, \bibinfo {author} {\bibfnamefont {Z.}~\bibnamefont {Liu}}, \ and\
  \bibinfo {author} {\bibfnamefont {Z.}~\bibnamefont {Fang}},\ }\href@noop {}
  {\bibfield  {journal} {\bibinfo  {journal} {ACS nano}\ }\textbf {\bibinfo
  {volume} {11}},\ \bibinfo {pages} {1165} (\bibinfo {year}
  {2017})}\BibitemShut {NoStop}%
\bibitem [{\citenamefont {Zeng}\ \emph {et~al.}(2017)\citenamefont {Zeng},
  \citenamefont {Li}, \citenamefont {Chen}, \citenamefont {Liao}, \citenamefont
  {Lou},\ and\ \citenamefont {Chen}}]{zeng2017highly}%
  \BibitemOpen
  \bibfield  {author} {\bibinfo {author} {\bibfnamefont {Y.}~\bibnamefont
  {Zeng}}, \bibinfo {author} {\bibfnamefont {X.}~\bibnamefont {Li}}, \bibinfo
  {author} {\bibfnamefont {W.}~\bibnamefont {Chen}}, \bibinfo {author}
  {\bibfnamefont {J.}~\bibnamefont {Liao}}, \bibinfo {author} {\bibfnamefont
  {J.}~\bibnamefont {Lou}}, \ and\ \bibinfo {author} {\bibfnamefont
  {Q.}~\bibnamefont {Chen}},\ }\href@noop {} {\bibfield  {journal} {\bibinfo
  {journal} {Advanced Materials Interfaces}\ }\textbf {\bibinfo {volume} {4}},\
  \bibinfo {pages} {1700739} (\bibinfo {year} {2017})}\BibitemShut {NoStop}%
\bibitem [{\citenamefont {Schaibley}\ \emph {et~al.}(2016)\citenamefont
  {Schaibley}, \citenamefont {Yu}, \citenamefont {Clark}, \citenamefont
  {Rivera}, \citenamefont {Ross}, \citenamefont {Seyler}, \citenamefont {Yao},\
  and\ \citenamefont {Xu}}]{schaibley2016valleytronics}%
  \BibitemOpen
  \bibfield  {author} {\bibinfo {author} {\bibfnamefont {J.~R.}\ \bibnamefont
  {Schaibley}}, \bibinfo {author} {\bibfnamefont {H.}~\bibnamefont {Yu}},
  \bibinfo {author} {\bibfnamefont {G.}~\bibnamefont {Clark}}, \bibinfo
  {author} {\bibfnamefont {P.}~\bibnamefont {Rivera}}, \bibinfo {author}
  {\bibfnamefont {J.~S.}\ \bibnamefont {Ross}}, \bibinfo {author}
  {\bibfnamefont {K.~L.}\ \bibnamefont {Seyler}}, \bibinfo {author}
  {\bibfnamefont {W.}~\bibnamefont {Yao}}, \ and\ \bibinfo {author}
  {\bibfnamefont {X.}~\bibnamefont {Xu}},\ }\href@noop {} {\bibfield  {journal}
  {\bibinfo  {journal} {Nature Reviews Materials}\ }\textbf {\bibinfo {volume}
  {1}},\ \bibinfo {pages} {1} (\bibinfo {year} {2016})}\BibitemShut {NoStop}%
\bibitem [{\citenamefont {Mak}\ \emph {et~al.}(2018)\citenamefont {Mak},
  \citenamefont {Xiao},\ and\ \citenamefont {Shan}}]{mak2018light}%
  \BibitemOpen
  \bibfield  {author} {\bibinfo {author} {\bibfnamefont {K.~F.}\ \bibnamefont
  {Mak}}, \bibinfo {author} {\bibfnamefont {D.}~\bibnamefont {Xiao}}, \ and\
  \bibinfo {author} {\bibfnamefont {J.}~\bibnamefont {Shan}},\ }\href@noop {}
  {\bibfield  {journal} {\bibinfo  {journal} {Nature Photonics}\ }\textbf
  {\bibinfo {volume} {12}},\ \bibinfo {pages} {451} (\bibinfo {year}
  {2018})}\BibitemShut {NoStop}%
\bibitem [{\citenamefont {Mak}\ \emph {et~al.}(2012)\citenamefont {Mak},
  \citenamefont {He}, \citenamefont {Shan},\ and\ \citenamefont
  {Heinz}}]{mak2012control}%
  \BibitemOpen
  \bibfield  {author} {\bibinfo {author} {\bibfnamefont {K.~F.}\ \bibnamefont
  {Mak}}, \bibinfo {author} {\bibfnamefont {K.}~\bibnamefont {He}}, \bibinfo
  {author} {\bibfnamefont {J.}~\bibnamefont {Shan}}, \ and\ \bibinfo {author}
  {\bibfnamefont {T.~F.}\ \bibnamefont {Heinz}},\ }\href@noop {} {\bibfield
  {journal} {\bibinfo  {journal} {Nature nanotechnology}\ }\textbf {\bibinfo
  {volume} {7}},\ \bibinfo {pages} {494} (\bibinfo {year} {2012})}\BibitemShut
  {NoStop}%
\bibitem [{\citenamefont {Splendiani}\ \emph {et~al.}(2010)\citenamefont
  {Splendiani}, \citenamefont {Sun}, \citenamefont {Zhang}, \citenamefont {Li},
  \citenamefont {Kim}, \citenamefont {Chim}, \citenamefont {Galli},\ and\
  \citenamefont {Wang}}]{splendiani2010emerging}%
  \BibitemOpen
  \bibfield  {author} {\bibinfo {author} {\bibfnamefont {A.}~\bibnamefont
  {Splendiani}}, \bibinfo {author} {\bibfnamefont {L.}~\bibnamefont {Sun}},
  \bibinfo {author} {\bibfnamefont {Y.}~\bibnamefont {Zhang}}, \bibinfo
  {author} {\bibfnamefont {T.}~\bibnamefont {Li}}, \bibinfo {author}
  {\bibfnamefont {J.}~\bibnamefont {Kim}}, \bibinfo {author} {\bibfnamefont
  {C.-Y.}\ \bibnamefont {Chim}}, \bibinfo {author} {\bibfnamefont
  {G.}~\bibnamefont {Galli}}, \ and\ \bibinfo {author} {\bibfnamefont
  {F.}~\bibnamefont {Wang}},\ }\href@noop {} {\bibfield  {journal} {\bibinfo
  {journal} {Nano letters}\ }\textbf {\bibinfo {volume} {10}},\ \bibinfo
  {pages} {1271} (\bibinfo {year} {2010})}\BibitemShut {NoStop}%
\bibitem [{\citenamefont {Zhu}\ \emph {et~al.}(2015)\citenamefont {Zhu},
  \citenamefont {Chen},\ and\ \citenamefont {Cui}}]{zhu2015exciton}%
  \BibitemOpen
  \bibfield  {author} {\bibinfo {author} {\bibfnamefont {B.}~\bibnamefont
  {Zhu}}, \bibinfo {author} {\bibfnamefont {X.}~\bibnamefont {Chen}}, \ and\
  \bibinfo {author} {\bibfnamefont {X.}~\bibnamefont {Cui}},\ }\href@noop {}
  {\bibfield  {journal} {\bibinfo  {journal} {Scientific reports}\ }\textbf
  {\bibinfo {volume} {5}},\ \bibinfo {pages} {1} (\bibinfo {year}
  {2015})}\BibitemShut {NoStop}%
\bibitem [{\citenamefont {Xiao}\ \emph {et~al.}(2017)\citenamefont {Xiao},
  \citenamefont {Zhao}, \citenamefont {Wang},\ and\ \citenamefont
  {Zhang}}]{xiao2017excitons}%
  \BibitemOpen
  \bibfield  {author} {\bibinfo {author} {\bibfnamefont {J.}~\bibnamefont
  {Xiao}}, \bibinfo {author} {\bibfnamefont {M.}~\bibnamefont {Zhao}}, \bibinfo
  {author} {\bibfnamefont {Y.}~\bibnamefont {Wang}}, \ and\ \bibinfo {author}
  {\bibfnamefont {X.}~\bibnamefont {Zhang}},\ }\href@noop {} {\bibfield
  {journal} {\bibinfo  {journal} {Nanophotonics}\ }\textbf {\bibinfo {volume}
  {6}},\ \bibinfo {pages} {1309} (\bibinfo {year} {2017})}\BibitemShut
  {NoStop}%
\bibitem [{\citenamefont {Amani}\ \emph {et~al.}(2015)\citenamefont {Amani},
  \citenamefont {Lien}, \citenamefont {Kiriya}, \citenamefont {Xiao},
  \citenamefont {Azcatl}, \citenamefont {Noh}, \citenamefont {Madhvapathy},
  \citenamefont {Addou}, \citenamefont {Kc}, \citenamefont {Dubey} \emph
  {et~al.}}]{amani2015near}%
  \BibitemOpen
  \bibfield  {author} {\bibinfo {author} {\bibfnamefont {M.}~\bibnamefont
  {Amani}}, \bibinfo {author} {\bibfnamefont {D.-H.}\ \bibnamefont {Lien}},
  \bibinfo {author} {\bibfnamefont {D.}~\bibnamefont {Kiriya}}, \bibinfo
  {author} {\bibfnamefont {J.}~\bibnamefont {Xiao}}, \bibinfo {author}
  {\bibfnamefont {A.}~\bibnamefont {Azcatl}}, \bibinfo {author} {\bibfnamefont
  {J.}~\bibnamefont {Noh}}, \bibinfo {author} {\bibfnamefont {S.~R.}\
  \bibnamefont {Madhvapathy}}, \bibinfo {author} {\bibfnamefont
  {R.}~\bibnamefont {Addou}}, \bibinfo {author} {\bibfnamefont
  {S.}~\bibnamefont {Kc}}, \bibinfo {author} {\bibfnamefont {M.}~\bibnamefont
  {Dubey}},  \emph {et~al.},\ }\href@noop {} {\bibfield  {journal} {\bibinfo
  {journal} {Science}\ }\textbf {\bibinfo {volume} {350}},\ \bibinfo {pages}
  {1065} (\bibinfo {year} {2015})}\BibitemShut {NoStop}%
\bibitem [{\citenamefont {Amani}\ \emph {et~al.}(2016)\citenamefont {Amani},
  \citenamefont {Burke}, \citenamefont {Ji}, \citenamefont {Zhao},
  \citenamefont {Lien}, \citenamefont {Taheri}, \citenamefont {Ahn},
  \citenamefont {Kirya}, \citenamefont {Ager~III}, \citenamefont {Yablonovitch}
  \emph {et~al.}}]{amani2016high}%
  \BibitemOpen
  \bibfield  {author} {\bibinfo {author} {\bibfnamefont {M.}~\bibnamefont
  {Amani}}, \bibinfo {author} {\bibfnamefont {R.~A.}\ \bibnamefont {Burke}},
  \bibinfo {author} {\bibfnamefont {X.}~\bibnamefont {Ji}}, \bibinfo {author}
  {\bibfnamefont {P.}~\bibnamefont {Zhao}}, \bibinfo {author} {\bibfnamefont
  {D.-H.}\ \bibnamefont {Lien}}, \bibinfo {author} {\bibfnamefont
  {P.}~\bibnamefont {Taheri}}, \bibinfo {author} {\bibfnamefont {G.~H.}\
  \bibnamefont {Ahn}}, \bibinfo {author} {\bibfnamefont {D.}~\bibnamefont
  {Kirya}}, \bibinfo {author} {\bibfnamefont {J.~W.}\ \bibnamefont {Ager~III}},
  \bibinfo {author} {\bibfnamefont {E.}~\bibnamefont {Yablonovitch}},  \emph
  {et~al.},\ }\href@noop {} {\bibfield  {journal} {\bibinfo  {journal} {ACS
  nano}\ }\textbf {\bibinfo {volume} {10}},\ \bibinfo {pages} {6535} (\bibinfo
  {year} {2016})}\BibitemShut {NoStop}%
\bibitem [{\citenamefont {Mouri}\ \emph {et~al.}(2013)\citenamefont {Mouri},
  \citenamefont {Miyauchi},\ and\ \citenamefont {Matsuda}}]{mouri2013tunable}%
  \BibitemOpen
  \bibfield  {author} {\bibinfo {author} {\bibfnamefont {S.}~\bibnamefont
  {Mouri}}, \bibinfo {author} {\bibfnamefont {Y.}~\bibnamefont {Miyauchi}}, \
  and\ \bibinfo {author} {\bibfnamefont {K.}~\bibnamefont {Matsuda}},\
  }\href@noop {} {\bibfield  {journal} {\bibinfo  {journal} {Nano letters}\
  }\textbf {\bibinfo {volume} {13}},\ \bibinfo {pages} {5944} (\bibinfo {year}
  {2013})}\BibitemShut {NoStop}%
\bibitem [{\citenamefont {Akselrod}\ \emph {et~al.}(2015)\citenamefont
  {Akselrod}, \citenamefont {Ming}, \citenamefont {Argyropoulos}, \citenamefont
  {Hoang}, \citenamefont {Lin}, \citenamefont {Ling}, \citenamefont {Smith},
  \citenamefont {Kong},\ and\ \citenamefont
  {Mikkelsen}}]{akselrod2015leveraging}%
  \BibitemOpen
  \bibfield  {author} {\bibinfo {author} {\bibfnamefont {G.~M.}\ \bibnamefont
  {Akselrod}}, \bibinfo {author} {\bibfnamefont {T.}~\bibnamefont {Ming}},
  \bibinfo {author} {\bibfnamefont {C.}~\bibnamefont {Argyropoulos}}, \bibinfo
  {author} {\bibfnamefont {T.~B.}\ \bibnamefont {Hoang}}, \bibinfo {author}
  {\bibfnamefont {Y.}~\bibnamefont {Lin}}, \bibinfo {author} {\bibfnamefont
  {X.}~\bibnamefont {Ling}}, \bibinfo {author} {\bibfnamefont {D.~R.}\
  \bibnamefont {Smith}}, \bibinfo {author} {\bibfnamefont {J.}~\bibnamefont
  {Kong}}, \ and\ \bibinfo {author} {\bibfnamefont {M.~H.}\ \bibnamefont
  {Mikkelsen}},\ }\href@noop {} {\bibfield  {journal} {\bibinfo  {journal}
  {Nano Letters}\ }\textbf {\bibinfo {volume} {15}},\ \bibinfo {pages} {3578}
  (\bibinfo {year} {2015})}\BibitemShut {NoStop}%
\bibitem [{\citenamefont {Wen}\ \emph {et~al.}(2017)\citenamefont {Wen},
  \citenamefont {Wang}, \citenamefont {Wang}, \citenamefont {Deng},
  \citenamefont {Zhuang}, \citenamefont {Zhang}, \citenamefont {Liu},
  \citenamefont {She}, \citenamefont {Chen}, \citenamefont {Chen} \emph
  {et~al.}}]{wen2017room}%
  \BibitemOpen
  \bibfield  {author} {\bibinfo {author} {\bibfnamefont {J.}~\bibnamefont
  {Wen}}, \bibinfo {author} {\bibfnamefont {H.}~\bibnamefont {Wang}}, \bibinfo
  {author} {\bibfnamefont {W.}~\bibnamefont {Wang}}, \bibinfo {author}
  {\bibfnamefont {Z.}~\bibnamefont {Deng}}, \bibinfo {author} {\bibfnamefont
  {C.}~\bibnamefont {Zhuang}}, \bibinfo {author} {\bibfnamefont
  {Y.}~\bibnamefont {Zhang}}, \bibinfo {author} {\bibfnamefont
  {F.}~\bibnamefont {Liu}}, \bibinfo {author} {\bibfnamefont {J.}~\bibnamefont
  {She}}, \bibinfo {author} {\bibfnamefont {J.}~\bibnamefont {Chen}}, \bibinfo
  {author} {\bibfnamefont {H.}~\bibnamefont {Chen}},  \emph {et~al.},\
  }\href@noop {} {\bibfield  {journal} {\bibinfo  {journal} {Nano letters}\
  }\textbf {\bibinfo {volume} {17}},\ \bibinfo {pages} {4689} (\bibinfo {year}
  {2017})}\BibitemShut {NoStop}%
\bibitem [{\citenamefont {Butun}\ \emph {et~al.}(2015)\citenamefont {Butun},
  \citenamefont {Tongay},\ and\ \citenamefont {Aydin}}]{butun2015enhanced}%
  \BibitemOpen
  \bibfield  {author} {\bibinfo {author} {\bibfnamefont {S.}~\bibnamefont
  {Butun}}, \bibinfo {author} {\bibfnamefont {S.}~\bibnamefont {Tongay}}, \
  and\ \bibinfo {author} {\bibfnamefont {K.}~\bibnamefont {Aydin}},\
  }\href@noop {} {\bibfield  {journal} {\bibinfo  {journal} {Nano letters}\
  }\textbf {\bibinfo {volume} {15}},\ \bibinfo {pages} {2700} (\bibinfo {year}
  {2015})}\BibitemShut {NoStop}%
\bibitem [{\citenamefont {You}\ \emph {et~al.}(2022)\citenamefont {You},
  \citenamefont {Li}, \citenamefont {Li}, \citenamefont {Qiu}, \citenamefont
  {Bi}, \citenamefont {Zhang}, \citenamefont {Zhang}, \citenamefont {Fang},\
  and\ \citenamefont {Wang}}]{you2022resonance}%
  \BibitemOpen
  \bibfield  {author} {\bibinfo {author} {\bibfnamefont {Q.}~\bibnamefont
  {You}}, \bibinfo {author} {\bibfnamefont {Z.}~\bibnamefont {Li}}, \bibinfo
  {author} {\bibfnamefont {Y.}~\bibnamefont {Li}}, \bibinfo {author}
  {\bibfnamefont {L.}~\bibnamefont {Qiu}}, \bibinfo {author} {\bibfnamefont
  {X.}~\bibnamefont {Bi}}, \bibinfo {author} {\bibfnamefont {L.}~\bibnamefont
  {Zhang}}, \bibinfo {author} {\bibfnamefont {D.}~\bibnamefont {Zhang}},
  \bibinfo {author} {\bibfnamefont {Y.}~\bibnamefont {Fang}}, \ and\ \bibinfo
  {author} {\bibfnamefont {P.}~\bibnamefont {Wang}},\ }\href@noop {} {\bibfield
   {journal} {\bibinfo  {journal} {ACS Applied Materials \& Interfaces}\
  }\textbf {\bibinfo {volume} {14}},\ \bibinfo {pages} {23756} (\bibinfo {year}
  {2022})}\BibitemShut {NoStop}%
\bibitem [{\citenamefont {Shinomiya}\ \emph {et~al.}(2022)\citenamefont
  {Shinomiya}, \citenamefont {Sugimoto}, \citenamefont {Hinamoto},
  \citenamefont {Lee}, \citenamefont {Brongersma},\ and\ \citenamefont
  {Fujii}}]{shinomiya2022enhanced}%
  \BibitemOpen
  \bibfield  {author} {\bibinfo {author} {\bibfnamefont {H.}~\bibnamefont
  {Shinomiya}}, \bibinfo {author} {\bibfnamefont {H.}~\bibnamefont {Sugimoto}},
  \bibinfo {author} {\bibfnamefont {T.}~\bibnamefont {Hinamoto}}, \bibinfo
  {author} {\bibfnamefont {Y.~J.}\ \bibnamefont {Lee}}, \bibinfo {author}
  {\bibfnamefont {M.~L.}\ \bibnamefont {Brongersma}}, \ and\ \bibinfo {author}
  {\bibfnamefont {M.}~\bibnamefont {Fujii}},\ }\href@noop {} {\bibfield
  {journal} {\bibinfo  {journal} {ACS Photonics}\ }\textbf {\bibinfo {volume}
  {9}},\ \bibinfo {pages} {1741} (\bibinfo {year} {2022})}\BibitemShut
  {NoStop}%
\bibitem [{\citenamefont {Pan}\ \emph {et~al.}(2022)\citenamefont {Pan},
  \citenamefont {Kang}, \citenamefont {Li}, \citenamefont {Zhang},
  \citenamefont {Li},\ and\ \citenamefont {Yang}}]{pan2022highly}%
  \BibitemOpen
  \bibfield  {author} {\bibinfo {author} {\bibfnamefont {R.}~\bibnamefont
  {Pan}}, \bibinfo {author} {\bibfnamefont {J.}~\bibnamefont {Kang}}, \bibinfo
  {author} {\bibfnamefont {Y.}~\bibnamefont {Li}}, \bibinfo {author}
  {\bibfnamefont {Z.}~\bibnamefont {Zhang}}, \bibinfo {author} {\bibfnamefont
  {R.}~\bibnamefont {Li}}, \ and\ \bibinfo {author} {\bibfnamefont
  {Y.}~\bibnamefont {Yang}},\ }\href@noop {} {\bibfield  {journal} {\bibinfo
  {journal} {ACS Applied Materials \& Interfaces}\ }\textbf {\bibinfo {volume}
  {14}},\ \bibinfo {pages} {12495} (\bibinfo {year} {2022})}\BibitemShut
  {NoStop}%
\bibitem [{\citenamefont {Ringler}\ \emph {et~al.}(2008)\citenamefont
  {Ringler}, \citenamefont {Schwemer}, \citenamefont {Wunderlich},
  \citenamefont {Nichtl}, \citenamefont {K{\"u}rzinger}, \citenamefont {Klar},\
  and\ \citenamefont {Feldmann}}]{ringler2008shaping}%
  \BibitemOpen
  \bibfield  {author} {\bibinfo {author} {\bibfnamefont {M.}~\bibnamefont
  {Ringler}}, \bibinfo {author} {\bibfnamefont {A.}~\bibnamefont {Schwemer}},
  \bibinfo {author} {\bibfnamefont {M.}~\bibnamefont {Wunderlich}}, \bibinfo
  {author} {\bibfnamefont {A.}~\bibnamefont {Nichtl}}, \bibinfo {author}
  {\bibfnamefont {K.}~\bibnamefont {K{\"u}rzinger}}, \bibinfo {author}
  {\bibfnamefont {T.}~\bibnamefont {Klar}}, \ and\ \bibinfo {author}
  {\bibfnamefont {J.}~\bibnamefont {Feldmann}},\ }\href@noop {} {\bibfield
  {journal} {\bibinfo  {journal} {Physical review letters}\ }\textbf {\bibinfo
  {volume} {100}},\ \bibinfo {pages} {203002} (\bibinfo {year}
  {2008})}\BibitemShut {NoStop}%
\bibitem [{\citenamefont {Li}\ \emph {et~al.}(2022)\citenamefont {Li},
  \citenamefont {Lu}, \citenamefont {Li}, \citenamefont {Shi}, \citenamefont
  {Yue},\ and\ \citenamefont {Zhao}}]{li2022plasmon}%
  \BibitemOpen
  \bibfield  {author} {\bibinfo {author} {\bibfnamefont {D.}~\bibnamefont
  {Li}}, \bibinfo {author} {\bibfnamefont {H.}~\bibnamefont {Lu}}, \bibinfo
  {author} {\bibfnamefont {Y.}~\bibnamefont {Li}}, \bibinfo {author}
  {\bibfnamefont {S.}~\bibnamefont {Shi}}, \bibinfo {author} {\bibfnamefont
  {Z.}~\bibnamefont {Yue}}, \ and\ \bibinfo {author} {\bibfnamefont
  {J.}~\bibnamefont {Zhao}},\ }\href@noop {} {\bibfield  {journal} {\bibinfo
  {journal} {Nanophotonics}\ }\textbf {\bibinfo {volume} {11}},\ \bibinfo
  {pages} {995} (\bibinfo {year} {2022})}\BibitemShut {NoStop}%
\bibitem [{\citenamefont {Sahoo}\ \emph {et~al.}(2023)\citenamefont {Sahoo},
  \citenamefont {Vs},\ and\ \citenamefont {Kumar}}]{sahoo2023electroplating}%
  \BibitemOpen
  \bibfield  {author} {\bibinfo {author} {\bibfnamefont {M.~K.}\ \bibnamefont
  {Sahoo}}, \bibinfo {author} {\bibfnamefont {A.~A.}\ \bibnamefont {Vs}}, \
  and\ \bibinfo {author} {\bibfnamefont {A.}~\bibnamefont {Kumar}},\
  }\href@noop {} {\bibfield  {journal} {\bibinfo  {journal} {Nanotechnology}\
  }\textbf {\bibinfo {volume} {34}},\ \bibinfo {pages} {195301} (\bibinfo
  {year} {2023})}\BibitemShut {NoStop}%
\bibitem [{\citenamefont {Mohapatra}\ \emph {et~al.}(2016)\citenamefont
  {Mohapatra}, \citenamefont {Deb}, \citenamefont {Singh}, \citenamefont
  {Vasa},\ and\ \citenamefont {Dhar}}]{mohapatra2016strictly}%
  \BibitemOpen
  \bibfield  {author} {\bibinfo {author} {\bibfnamefont {P.}~\bibnamefont
  {Mohapatra}}, \bibinfo {author} {\bibfnamefont {S.}~\bibnamefont {Deb}},
  \bibinfo {author} {\bibfnamefont {B.}~\bibnamefont {Singh}}, \bibinfo
  {author} {\bibfnamefont {P.}~\bibnamefont {Vasa}}, \ and\ \bibinfo {author}
  {\bibfnamefont {S.}~\bibnamefont {Dhar}},\ }\href@noop {} {\bibfield
  {journal} {\bibinfo  {journal} {Applied Physics Letters}\ }\textbf {\bibinfo
  {volume} {108}},\ \bibinfo {pages} {042101} (\bibinfo {year}
  {2016})}\BibitemShut {NoStop}%
\bibitem [{\citenamefont {Gurarslan}\ \emph {et~al.}(2014)\citenamefont
  {Gurarslan}, \citenamefont {Yu}, \citenamefont {Su}, \citenamefont {Yu},
  \citenamefont {Suarez}, \citenamefont {Yao}, \citenamefont {Zhu},
  \citenamefont {Ozturk}, \citenamefont {Zhang},\ and\ \citenamefont
  {Cao}}]{gurarslan2014surface}%
  \BibitemOpen
  \bibfield  {author} {\bibinfo {author} {\bibfnamefont {A.}~\bibnamefont
  {Gurarslan}}, \bibinfo {author} {\bibfnamefont {Y.}~\bibnamefont {Yu}},
  \bibinfo {author} {\bibfnamefont {L.}~\bibnamefont {Su}}, \bibinfo {author}
  {\bibfnamefont {Y.}~\bibnamefont {Yu}}, \bibinfo {author} {\bibfnamefont
  {F.}~\bibnamefont {Suarez}}, \bibinfo {author} {\bibfnamefont
  {S.}~\bibnamefont {Yao}}, \bibinfo {author} {\bibfnamefont {Y.}~\bibnamefont
  {Zhu}}, \bibinfo {author} {\bibfnamefont {M.}~\bibnamefont {Ozturk}},
  \bibinfo {author} {\bibfnamefont {Y.}~\bibnamefont {Zhang}}, \ and\ \bibinfo
  {author} {\bibfnamefont {L.}~\bibnamefont {Cao}},\ }\href@noop {} {\bibfield
  {journal} {\bibinfo  {journal} {ACS nano}\ }\textbf {\bibinfo {volume} {8}},\
  \bibinfo {pages} {11522} (\bibinfo {year} {2014})}\BibitemShut {NoStop}%
\bibitem [{\citenamefont {Palik}(1998)}]{palik1998handbook}%
  \BibitemOpen
  \bibfield  {author} {\bibinfo {author} {\bibfnamefont {E.~D.}\ \bibnamefont
  {Palik}},\ }\href@noop {} {\emph {\bibinfo {title} {Handbook of optical
  constants of solids}}},\ Vol.~\bibinfo {volume} {3}\ (\bibinfo  {publisher}
  {Academic press},\ \bibinfo {year} {1998})\BibitemShut {NoStop}%
\bibitem [{\citenamefont {Li}\ \emph {et~al.}(2012)\citenamefont {Li},
  \citenamefont {Zhang}, \citenamefont {Yap}, \citenamefont {Tay},
  \citenamefont {Edwin}, \citenamefont {Olivier},\ and\ \citenamefont
  {Baillargeat}}]{li2012bulk}%
  \BibitemOpen
  \bibfield  {author} {\bibinfo {author} {\bibfnamefont {H.}~\bibnamefont
  {Li}}, \bibinfo {author} {\bibfnamefont {Q.}~\bibnamefont {Zhang}}, \bibinfo
  {author} {\bibfnamefont {C.~C.~R.}\ \bibnamefont {Yap}}, \bibinfo {author}
  {\bibfnamefont {B.~K.}\ \bibnamefont {Tay}}, \bibinfo {author} {\bibfnamefont
  {T.~H.~T.}\ \bibnamefont {Edwin}}, \bibinfo {author} {\bibfnamefont
  {A.}~\bibnamefont {Olivier}}, \ and\ \bibinfo {author} {\bibfnamefont
  {D.}~\bibnamefont {Baillargeat}},\ }\href@noop {} {\bibfield  {journal}
  {\bibinfo  {journal} {Advanced Functional Materials}\ }\textbf {\bibinfo
  {volume} {22}},\ \bibinfo {pages} {1385} (\bibinfo {year}
  {2012})}\BibitemShut {NoStop}%
\bibitem [{\citenamefont {Plechinger}\ \emph {et~al.}(2014)\citenamefont
  {Plechinger}, \citenamefont {Mann}, \citenamefont {Preciado}, \citenamefont
  {Barroso}, \citenamefont {Nguyen}, \citenamefont {Eroms}, \citenamefont
  {Sch\"{u}ller}, \citenamefont {Bartels},\ and\ \citenamefont
  {Korn}}]{Plechinger2014}%
  \BibitemOpen
  \bibfield  {author} {\bibinfo {author} {\bibfnamefont {G.}~\bibnamefont
  {Plechinger}}, \bibinfo {author} {\bibfnamefont {J.}~\bibnamefont {Mann}},
  \bibinfo {author} {\bibfnamefont {E.}~\bibnamefont {Preciado}}, \bibinfo
  {author} {\bibfnamefont {D.}~\bibnamefont {Barroso}}, \bibinfo {author}
  {\bibfnamefont {A.}~\bibnamefont {Nguyen}}, \bibinfo {author} {\bibfnamefont
  {J.}~\bibnamefont {Eroms}}, \bibinfo {author} {\bibfnamefont
  {C.}~\bibnamefont {Sch\"{u}ller}}, \bibinfo {author} {\bibfnamefont
  {L.}~\bibnamefont {Bartels}}, \ and\ \bibinfo {author} {\bibfnamefont
  {T.}~\bibnamefont {Korn}},\ }\href {\doibase 10.1088/0268-1242/29/6/064008}
  {\bibfield  {journal} {\bibinfo  {journal} {Semiconductor Science and
  Technology}\ }\textbf {\bibinfo {volume} {29}},\ \bibinfo {pages} {064008}
  (\bibinfo {year} {2014})}\BibitemShut {NoStop}%
\bibitem [{\citenamefont {Bhanu}\ \emph {et~al.}(2014)\citenamefont {Bhanu},
  \citenamefont {Islam}, \citenamefont {Tetard},\ and\ \citenamefont
  {Khondaker}}]{bhanu2014photoluminescence}%
  \BibitemOpen
  \bibfield  {author} {\bibinfo {author} {\bibfnamefont {U.}~\bibnamefont
  {Bhanu}}, \bibinfo {author} {\bibfnamefont {M.~R.}\ \bibnamefont {Islam}},
  \bibinfo {author} {\bibfnamefont {L.}~\bibnamefont {Tetard}}, \ and\ \bibinfo
  {author} {\bibfnamefont {S.~I.}\ \bibnamefont {Khondaker}},\ }\href@noop {}
  {\bibfield  {journal} {\bibinfo  {journal} {Scientific reports}\ }\textbf
  {\bibinfo {volume} {4}},\ \bibinfo {pages} {5575} (\bibinfo {year}
  {2014})}\BibitemShut {NoStop}%
\bibitem [{\citenamefont {Dumcenco}\ \emph {et~al.}(2015)\citenamefont
  {Dumcenco}, \citenamefont {Ovchinnikov}, \citenamefont {Sanchez},
  \citenamefont {Gillet}, \citenamefont {Alexander}, \citenamefont {Lazar},
  \citenamefont {Radenovic},\ and\ \citenamefont {Kis}}]{dumcenco2015large}%
  \BibitemOpen
  \bibfield  {author} {\bibinfo {author} {\bibfnamefont {D.}~\bibnamefont
  {Dumcenco}}, \bibinfo {author} {\bibfnamefont {D.}~\bibnamefont
  {Ovchinnikov}}, \bibinfo {author} {\bibfnamefont {O.~L.}\ \bibnamefont
  {Sanchez}}, \bibinfo {author} {\bibfnamefont {P.}~\bibnamefont {Gillet}},
  \bibinfo {author} {\bibfnamefont {D.~T.}\ \bibnamefont {Alexander}}, \bibinfo
  {author} {\bibfnamefont {S.}~\bibnamefont {Lazar}}, \bibinfo {author}
  {\bibfnamefont {A.}~\bibnamefont {Radenovic}}, \ and\ \bibinfo {author}
  {\bibfnamefont {A.}~\bibnamefont {Kis}},\ }\href@noop {} {\bibfield
  {journal} {\bibinfo  {journal} {2D Materials}\ }\textbf {\bibinfo {volume}
  {2}},\ \bibinfo {pages} {044005} (\bibinfo {year} {2015})}\BibitemShut
  {NoStop}%
\bibitem [{\citenamefont {Kudryashov}\ \emph {et~al.}(1999)\citenamefont
  {Kudryashov}, \citenamefont {Mamakin},\ and\ \citenamefont
  {Yunovich}}]{kudryashov1999influence}%
  \BibitemOpen
  \bibfield  {author} {\bibinfo {author} {\bibfnamefont {V.}~\bibnamefont
  {Kudryashov}}, \bibinfo {author} {\bibfnamefont {S.}~\bibnamefont {Mamakin}},
  \ and\ \bibinfo {author} {\bibfnamefont {A.}~\bibnamefont {Yunovich}},\
  }\href@noop {} {\bibfield  {journal} {\bibinfo  {journal} {Technical Physics
  Letters}\ }\textbf {\bibinfo {volume} {25}} (\bibinfo {year}
  {1999})}\BibitemShut {NoStop}%
\bibitem [{\citenamefont {Wang}\ \emph {et~al.}(2016)\citenamefont {Wang},
  \citenamefont {Dong}, \citenamefont {Gu}, \citenamefont {Chang},
  \citenamefont {Zhang}, \citenamefont {Li}, \citenamefont {Zhao},
  \citenamefont {Eda}, \citenamefont {Zhang}, \citenamefont {Grinblat} \emph
  {et~al.}}]{wang2016giant}%
  \BibitemOpen
  \bibfield  {author} {\bibinfo {author} {\bibfnamefont {Z.}~\bibnamefont
  {Wang}}, \bibinfo {author} {\bibfnamefont {Z.}~\bibnamefont {Dong}}, \bibinfo
  {author} {\bibfnamefont {Y.}~\bibnamefont {Gu}}, \bibinfo {author}
  {\bibfnamefont {Y.-H.}\ \bibnamefont {Chang}}, \bibinfo {author}
  {\bibfnamefont {L.}~\bibnamefont {Zhang}}, \bibinfo {author} {\bibfnamefont
  {L.-J.}\ \bibnamefont {Li}}, \bibinfo {author} {\bibfnamefont
  {W.}~\bibnamefont {Zhao}}, \bibinfo {author} {\bibfnamefont {G.}~\bibnamefont
  {Eda}}, \bibinfo {author} {\bibfnamefont {W.}~\bibnamefont {Zhang}}, \bibinfo
  {author} {\bibfnamefont {G.}~\bibnamefont {Grinblat}},  \emph {et~al.},\
  }\href@noop {} {\bibfield  {journal} {\bibinfo  {journal} {Nature
  communications}\ }\textbf {\bibinfo {volume} {7}},\ \bibinfo {pages} {11283}
  (\bibinfo {year} {2016})}\BibitemShut {NoStop}%
\bibitem [{\citenamefont {Bharadwaj}\ and\ \citenamefont
  {Novotny}(2007)}]{Bharadwaj2007}%
  \BibitemOpen
  \bibfield  {author} {\bibinfo {author} {\bibfnamefont {P.}~\bibnamefont
  {Bharadwaj}}\ and\ \bibinfo {author} {\bibfnamefont {L.}~\bibnamefont
  {Novotny}},\ }\href {\doibase 10.1364/oe.15.014266} {\bibfield  {journal}
  {\bibinfo  {journal} {Optics Express}\ }\textbf {\bibinfo {volume} {15}},\
  \bibinfo {pages} {14266} (\bibinfo {year} {2007})}\BibitemShut {NoStop}%
\bibitem [{\citenamefont {Lien}\ \emph {et~al.}(2019)\citenamefont {Lien},
  \citenamefont {Uddin}, \citenamefont {Yeh}, \citenamefont {Amani},
  \citenamefont {Kim}, \citenamefont {Ager}, \citenamefont {Yablonovitch},\
  and\ \citenamefont {Javey}}]{Lien2019}%
  \BibitemOpen
  \bibfield  {author} {\bibinfo {author} {\bibfnamefont {D.-H.}\ \bibnamefont
  {Lien}}, \bibinfo {author} {\bibfnamefont {S.~Z.}\ \bibnamefont {Uddin}},
  \bibinfo {author} {\bibfnamefont {M.}~\bibnamefont {Yeh}}, \bibinfo {author}
  {\bibfnamefont {M.}~\bibnamefont {Amani}}, \bibinfo {author} {\bibfnamefont
  {H.}~\bibnamefont {Kim}}, \bibinfo {author} {\bibfnamefont {J.~W.}\
  \bibnamefont {Ager}}, \bibinfo {author} {\bibfnamefont {E.}~\bibnamefont
  {Yablonovitch}}, \ and\ \bibinfo {author} {\bibfnamefont {A.}~\bibnamefont
  {Javey}},\ }\href {\doibase 10.1126/science.aaw8053} {\bibfield  {journal}
  {\bibinfo  {journal} {Science}\ }\textbf {\bibinfo {volume} {364}},\ \bibinfo
  {pages} {468} (\bibinfo {year} {2019})}\BibitemShut {NoStop}%
\bibitem [{\citenamefont {Singh}\ \emph {et~al.}(2023)\citenamefont {Singh},
  \citenamefont {Mandal}, \citenamefont {Gupta}, \citenamefont {V.S.},
  \citenamefont {Eswaramoorthy}, \citenamefont {Kumar}, \citenamefont {Kala},
  \citenamefont {Dixit}, \citenamefont {Achanta},\ and\ \citenamefont
  {Kumar}}]{Singh2023}%
  \BibitemOpen
  \bibfield  {author} {\bibinfo {author} {\bibfnamefont {A.~K.}\ \bibnamefont
  {Singh}}, \bibinfo {author} {\bibfnamefont {K.~K.}\ \bibnamefont {Mandal}},
  \bibinfo {author} {\bibfnamefont {Y.}~\bibnamefont {Gupta}}, \bibinfo
  {author} {\bibfnamefont {A.~A.}\ \bibnamefont {V.S.}}, \bibinfo {author}
  {\bibfnamefont {L.}~\bibnamefont {Eswaramoorthy}}, \bibinfo {author}
  {\bibfnamefont {B.}~\bibnamefont {Kumar}}, \bibinfo {author} {\bibfnamefont
  {A.}~\bibnamefont {Kala}}, \bibinfo {author} {\bibfnamefont {S.}~\bibnamefont
  {Dixit}}, \bibinfo {author} {\bibfnamefont {V.~G.}\ \bibnamefont {Achanta}},
  \ and\ \bibinfo {author} {\bibfnamefont {A.}~\bibnamefont {Kumar}},\ }\href
  {\doibase 10.1103/physrevapplied.19.044012} {\bibfield  {journal} {\bibinfo
  {journal} {Physical Review Applied}\ }\textbf {\bibinfo {volume} {19}}
  (\bibinfo {year} {2023}),\ 10.1103/physrevapplied.19.044012}\BibitemShut
  {NoStop}%
\end{thebibliography}%

\noindent

\section*{Acknowledgements}

A.K. acknowledges funding support from the Department of Science and Technology via the grants: SB/S2/RJN-110/2017, ECR/2018/001485, and DST/NM/NS-2018/49. A.A.V.S  acknowledges UGC for his junior research fellowship to support his PhD. The authors acknowledge the Centre of Excellence in Nanoelectronics (CEN) at IIT Bombay for providing fabrication and characterisation facilities. M.K.S. acknowledges the Institute Postdoctoral fellowship for financial support. We thank Sandip Ghosh from TIFR Mumbai for help with some verification measurements.

\end{document}